\documentclass[conference]{IEEEtran}
\IEEEoverridecommandlockouts
\usepackage{cite}
\usepackage{amsmath,amssymb,amsfonts}
\usepackage{algorithmic}
\usepackage{graphicx}
\usepackage{textcomp}
\usepackage{xcolor}
\def\BibTeX{{\rm B\kern-.05em{\sc i\kern-.025em b}\kern-.08em
		T\kern-.1667em\lower.7ex\hbox{E}\kern-.125emX}}
\usepackage{bm}
\usepackage{subfigure}
\usepackage{graphicx,epsfig,balance}
\usepackage{multirow}
\usepackage{amsfonts,amssymb}
\usepackage{booktabs}
\usepackage{threeparttable}
\usepackage{soul}
\usepackage{array}
\usepackage{makecell}
\usepackage[linesnumbered,ruled,lined]{algorithm2e}

\usepackage{verbatim}
\begin{document}

\title{A Novel Channel Identification  Architecture for mmWave Systems Based on Eigen Features}

\author{Yibin Zhang$^\dag$, Jinlong Sun$^\dag$, Guan Gui$^\dag$, Haris Gacanin$^{\ddag}$ and Fumiyuki Adachi$^{\dag\dag}$\\ \\
$^\dag$College of Telecommunications and Information Engineering, NJUPT, Nanjing, China\\
$^{*}$Institute for Communication Technologies and Embedded Systems, RWTH Aachen University, Aachen, Germany\\
$^{\dag\dag}$Research Organization of Electrical Communication, Tohoku University, Sendai, Japan\\
}

\maketitle

\begin{abstract}
Millimeter wave (mmWave) communication technique has been developed rapidly because of many advantages of high speed, large bandwidth, and ultra-low delay. However, mmWave communications systems suffer from fast fading and frequent blocking. Hence, the ideal communication environment for mmWave is  line of sight (LOS) channel.  To improve the  efficiency and capacity of mmWave system, and to better build the Internet of Everything (IoE) service network, this paper focuses on the channel identification technique in line-of-sight (LOS) and non-LOS (NLOS) environments.
Considering the limited computing ability of user equipments (UEs), this paper proposes a novel channel identification architecture based on eigen features, i.e. eigenmatrix and eigenvector (EMEV) of channel state information (CSI). 
Furthermore, this paper explores clustered delay line (CDL) channel identification with mmWave, which is defined by the 3rd generation partnership project (3GPP). 
The experimental results show that the EMEV based scheme can achieve  identification accuracy of $99.88\%$ assuming perfect CSI. In the robustness test, the maximum noise  can be tolerated is SNR$=16$ dB, with the threshold $acc \geq 95\%$. What is more, the novel architecture based on EMEV feature will reduce the  comprehensive overhead by about $90\%$.
\end{abstract}


\section{Introduction}
With the mature applications of internet of things (IoT) communication \cite{LiYe_IoT}, the fifth generation (5G) and sixth generation (6G) \cite{Kato_6G1} of mobile communication has put forward new development vision: internet of everything (IoE).  The explosive network connection and data transmission demands  put forward higher wireless data traffic, e.g., a 1000-fold capacity increase \cite{survey1}. However, the microwave band (300 MHz to 3 GHz) cannot support the escalating capacity demand. Thus, exploring new spectrum with broader bandwidths, such as the millimeter wave (mmWave) bands (30--300 GHz), is a promising solution to increase network capacity \cite{survey2}. Intelligent wireless communication systems with mmWave band and massive multiple-input multiple-output (MIMO)  should  be  the key technology of future IoE network \cite{LiYe2019,Kato_6G2,Guojiajia}.  However, T. Mantoro \emph{et al.} \cite{Mantoro2017} claimed that the transmission performance of line of sight (LOS) and none line of sight (NLOS) with mmWave band is very different. This is caused by the coupling of time delay, received power, azimut angle of departure (AoD), elevation AoD, azimut angle of arrival (AoA) , elevation AoA, path-lost and RMS delay in LOS and NLOS environment.

In recent years, many scholars have been studying how to identify LOS and NLOS channels.
J. Zhang \emph{et al}. \cite{Zhang2013} explored the scheme of NLOS channel identification using kurtosis to improve the accuracy of indoor wireless localization problem. Meanwhile, C.X. Huang \emph{et al.} \cite{Huang2017} proposed a LOS-NLOS identification algorithm for indoor localization problem. R. Diamant \emph{et al.} \cite{Diamant2014} focused on the identification algorithm of LOS and NLOS in underwater communication environment. All papers \cite{Zhang2013, Huang2017, Diamant2014} used traditional algorithm to classify channels. They paid more attention to analyzing the channel state information (CSI)  to obtain the difference between LOS and NLOS channel. With the development of machine learning (ML) and deep learning (DL), more researchers \cite{Xiao2017,Zeng2018,Zhang2019,Huang2020,Jiang2020,Cui2021,Zheng2020} prefered to applying ML/DL algorithms to identify LOS and NLOS channel. X. Fu  \emph{et al.}  \cite{Xiao2017} proposed a real-time LOS/NLOS identification based on CSI characteristics and K-means algorithm. In this paper, the author claimed that they achieved great identification performance for both static and dynamic scenarios.  T.Y. Zeng \emph{et al.} \cite{Zeng2018}  applied convolutional neural network (CNN) to channel identification of three dimensional massive MIMO system. And the authors used channel model 3D urban macro (UMa) defined by the 3rd generation partnership project (3GPP) to verify the proposed scheme.  Aiming at the NLOS channel filtering problem in radio frequency identification (RFID), S.G. Zhang \emph{et al.}  \cite{Zhang2019} proposed a variety of efficient and novel algorithms. These methods included a new metric that combined both phase and received signal strength variances  and a ML based algorithm. C. Huang \emph{et al.} \cite{Huang2020} proposed a time-varying angular information-based LOS identification solution based on ML.  For indoor ultra-wideband (UWB) positioning systems, C.H. Jiang \emph{et al.} \cite{Jiang2020} and Z. Cui \emph{et al.} \cite{Cui2021} employed CNN to identify the NLOS signal.

To ensure the efficient transmission of mmWave systems, LOS/NLOS channel identification is necessary.  Although many works have proposed identification solutions for LOS/NLOS channels, they focused on the binary classification (0/1 classification) problem. So, this paper aims to explore a more accurate channel identification scheme.  3GPP report \cite{3GPP38901} introduced clustered delay line (CDL) channel model which is  defined for the full frequency range from 0.5 GHz to 100 GHz. CDL channel models can be further divided into two categories: NLOS and LOS, and five sub-categories: CDL-A, CDL-B, CDL-C, CDL-D and CDL-E.  In this paper, we propose a novel channel identification  architecture for  mmWave  systems. The main contributions of this paper are summarized below.
\begin{itemize}
\item This paper further explores the channel identification problem.  A  more exact CDL channel identification scheme is proposed, which can further identify the channel type among CDL-A to CDL-E. Accurate identification of channel type can help to improve efficiency and capacity for mmWave wireless communication system.
\item To avoid increasing the overhead of UEs, this paper intends to utilize eigenmatrix and eigenvector as identification obeject, instead of CSI matrix. Compared with traditional DL algorithm,  a lightweight channel identification architecture is designed for UEs.
\end{itemize}

\section{System Model and Problem Formulation}
This section is composed of the following two parts. First of all, we show the problem formulation and some channel knowledge to the readers, to better understand the purpose of this paper. Then, we introduce singular value decomposition (SVD) and the application of eigenvector and eigenmatrix.

\subsection{Channel Model and Problem Formulation}
This paper focuses on the identification solution for CDL channel model.  In order to better understand CDL channel model, this section will give simple channel modeling and LOS channel probability distribution function defined by 3GPP.  Considering a three-dimensional mmWave channel \cite{3GPP36873}, the CSI estimated at the UE can be expressed as (\ref{channel1}). It is not difficult to find  that the fading rate is inversely proportional to the  wavelength $\lambda$. In other words, mmWave wireless communication system will suffer faster fading as the wave length increases.
\begin{equation}
	\begin{aligned}
		h_{u,s,t}(t) & =  \sum_{m=1}^{M}\sqrt{P_{n,m}} [ c_{u,s,n,m} \cdot \exp(j2\pi v_{n,m}t) \\
		&  \cdot \exp\left(j2\pi \lambda^{-1} \mathbf{d}_s \phi_{n,m}\right) \cdot \exp(j2\pi \lambda^{-1} \mathbf{d}_s \phi^{'}_{n,m})]
	\end{aligned}
\label{channel1}
\end{equation}
where $P_{m,n}$ represents the power of ray $m$  in the ray cluster $n$. And the $c_{u,s,n,m}$ is the coefficient calculated by field patterns and initial random phases for a pair of antenna elements between BS $s$ and UE $u$. Then, $\mathbf{d}_s$ and $\mathbf{H}_u$ are the locations of the BS and UE, respectively, $\lambda$ is the wavelength, $\phi_{n,m}$ and $\phi^{'}_{n,m}$ represents the angle vectors of departure and arrival, $v_{n,m}$ denotes the speed and can be understood as Doppler shift parameter.

After channel model, we will further discuss the probability distribution of LOS channel.  Considering an urban macro (UMa) scenario defined by 3GPP TR38.901 \cite{3GPP38901}, we assume that the plane straight-line distance from the UE to the BS is $d_{2D}$ and the LOS probability is $\mathrm{Pr}_{LOS}$. If  $d_{2D} \leq 18 ~{\rm m}$, then $\mathrm{Pr}_{LOS} = 1$, else the $\mathrm{Pr}_{LOS}$ can be calculated via
\begin{equation}
	\begin{aligned}
		\mathrm{Pr}_{LOS}  =&  \left[ \frac{18}{d_{2D}}+\exp\left(-\frac{d_{2D}}{63}\right)\left(1-\frac{18}{d_{2D}}\right) \right] \\
			& \cdot \left[    1+ 0.8 \cdot C(h_{UT}) \left(\frac{d_{2D}}{100}\right)^3 \exp\left(-\frac{d_{2D}}{150}\right) \right]
	\end{aligned}
\label{PrLOS}
\end{equation}
where the $C(h_{UT})$ can be found in (\ref{chut}), and the $h_{UT}$ denotes the antenna height for the UE.
\begin{equation}
	C(h_{UT})=\left\{
	\begin{aligned}
		0 & , & h_{UT} \leq 13 ~{\rm m} \\
		\left(\frac{h_{UT-13}}{10}\right)^{1.5} & , & 13 ~{\rm m} \leq h_{UT} \leq 28 ~{\rm m}
	\end{aligned}
	\right.
\label{chut}
\end{equation}

In summary, NLOS channel  will be a more common scenario with the popularization of mmWave systems. In  fact, the position of the UE relative to BS is always changing, so  UEs need to frequently identify the channel type and reports to BS. Accurate channel type will help establish a more efficient and intelligent communication link between the UE and BS.

\subsection{Application of SVD transformation}
This section will show the advantages of SVD transformation and its application in wireless communication.  In order to reduce the conflict between multi-users and increase the channel capacity in MIMO channels, the transmitter needs to use the beamforming technology to precode the  data flow according to the quality of channel.  A classical precoding matrix is based on SVD transformation of CSI matrix.

Considering a CDL channel with $N_t$ BS antennas and $N_r $ UE antennas.  For simplicity, it is assumed that the number of RB  is 1, i.e. $N_{RB} = 1$. So, we can obtain our CSI matrix as $\mathbf{H} \in \mathbb{C}^{N_r \times N_t}$.  First, the CSI matrix $\mathbf{H}$ should carry on SVD transformation as
\begin{equation}
	\mathbf{H} = \mathbf{U} \cdot \mathbf{\Sigma} \cdot \mathbf{V}^*
	\label{svd}
\end{equation}
where $\mathbf{U} \in \mathbb{C}^{N_r \times N_r}$ and $\mathbf{V} \in \mathbb{C}^{N_t \times N_t}$ are the left-singular and the right-singular matrices\footnote{Both $\mathbf{U}$ and   $\mathbf{V}$ will be called as eigenmatrix in the follows.}, respectively. What is more,  $\mathbf{U}\mathbf{U}^*=\mathbf{I}_{N_r}, \mathbf{V}\mathbf{V}^*=\mathbf{I}_{N_t}$ \footnote{$\mathbf{X}^{*}$ denotes conjugate transpose matrix of $\mathbf{X}$.}. Note that $\mathbf{\Sigma} = (\Lambda,0)$ and $\Lambda$ can be expressed as follows:
\begin{equation}
	\Lambda = \left(
	\begin{array}{ccc}
		\sqrt{\lambda_1} & \cdots & 0 \\
		\vdots & \ddots & \vdots \\
		0 & \cdots & \sqrt{\lambda_{N_r}}
	\end{array}
	\right)_{N_r \times N_r}
\label{lambda}
\end{equation}
which represents the singular value matrix. And we define the eigenvalues of $\mathbf{H}\mathbf{H}^{*}$ as  $\mathbf{s} = \left[ \lambda_1, \lambda_2, \dots, \lambda_{N_r}   \right]$.

Next, the application of SVD transformation will be introduced in detail. The unitary  matrices $\mathbf{V}$ and   $\mathbf{U}$ are used as precoding matrix for transmitter and receiver, respectively.  When BS needs to sent the parallel data flow $\boldsymbol{x} = [x_1, x_2, \dots, x_{N_t}]^T$ to multiuser, right-singular matrix $\mathbf{V}$ will be used for precoding: $\boldsymbol{x}_t = \mathbf{V} \cdot \boldsymbol{x}$. Thirdly, we consider a classical signal transmission model as
\begin{equation}
	\boldsymbol{y} = \mathbf{H}\boldsymbol{x}_t + \boldsymbol{n}
	\label{channel}
\end{equation}
where $\boldsymbol{y}$ is the received data flow and $\boldsymbol{n}$ denotes the noise vector. The channel matrix $\mathbf{H}$  can be expressed by (\ref{svd}), and we can get
\begin{equation}
	\begin{aligned}
			\boldsymbol{y} & = \mathbf{U}  \mathbf{\Sigma} \mathbf{V}^*  \mathbf{V} \cdot \boldsymbol{x}+ \boldsymbol{n} \\
			& =  \mathbf{U}  \mathbf{\Sigma} \cdot \boldsymbol{x} + \boldsymbol{n}
	\end{aligned}
	\label{channel2}
\end{equation}
Finally, the receiver will use   $\mathbf{U}^*$ for de-precoding, which can be expressed as
\begin{equation}
	\begin{aligned}
		\mathbf{U}^{*} \boldsymbol{y} & =  \mathbf{U}^{*} (\mathbf{U}  \mathbf{\Sigma} \cdot \boldsymbol{x}+\boldsymbol{n})  \\
		& =  \mathbf{\Sigma} \boldsymbol{x} + \mathbf{U}^* \boldsymbol{n}
	\end{aligned}
	\label{depreco}
\end{equation}
The noise component in (\ref{depreco})  will be  filtered out by the receiver. So, the receiver can recover the data flow $\boldsymbol{x}$ by $\mathbf{\Sigma}$.

In short, as a receiver, UE should pay more attention to  $\mathbf{U}$ and  $\mathbf{S}$. While  used for the precoding algorithms for UEs, eigenvector and eigenmatrix are also the representation of the channel features in the channel matrix $\mathbf{H}$. Therefore, this paper considers using eigenvector and eigenmatrix to support UEs for channel identification.  The  scheme and the algorithm framework are described in section \uppercase\expandafter{\romannumeral3}.

\section{Proposed Channel Identification Architecture}
In this section, the proposed channel identification architecture based on eigenmatrix and eigenvector is presented. This section will be introduced according to the following three points. First, the  overall framework  of the proposed novel channel identification architecture will be shown. Then, we will give a detailed algorithm describtion. Finally,  the hyper-parameters and  overhead analysis of proposed  neural network will be given in detail.
\begin{figure}[htbp]
	\centering
	\includegraphics[width=3.5 in,trim=0 0 0 0,clip]{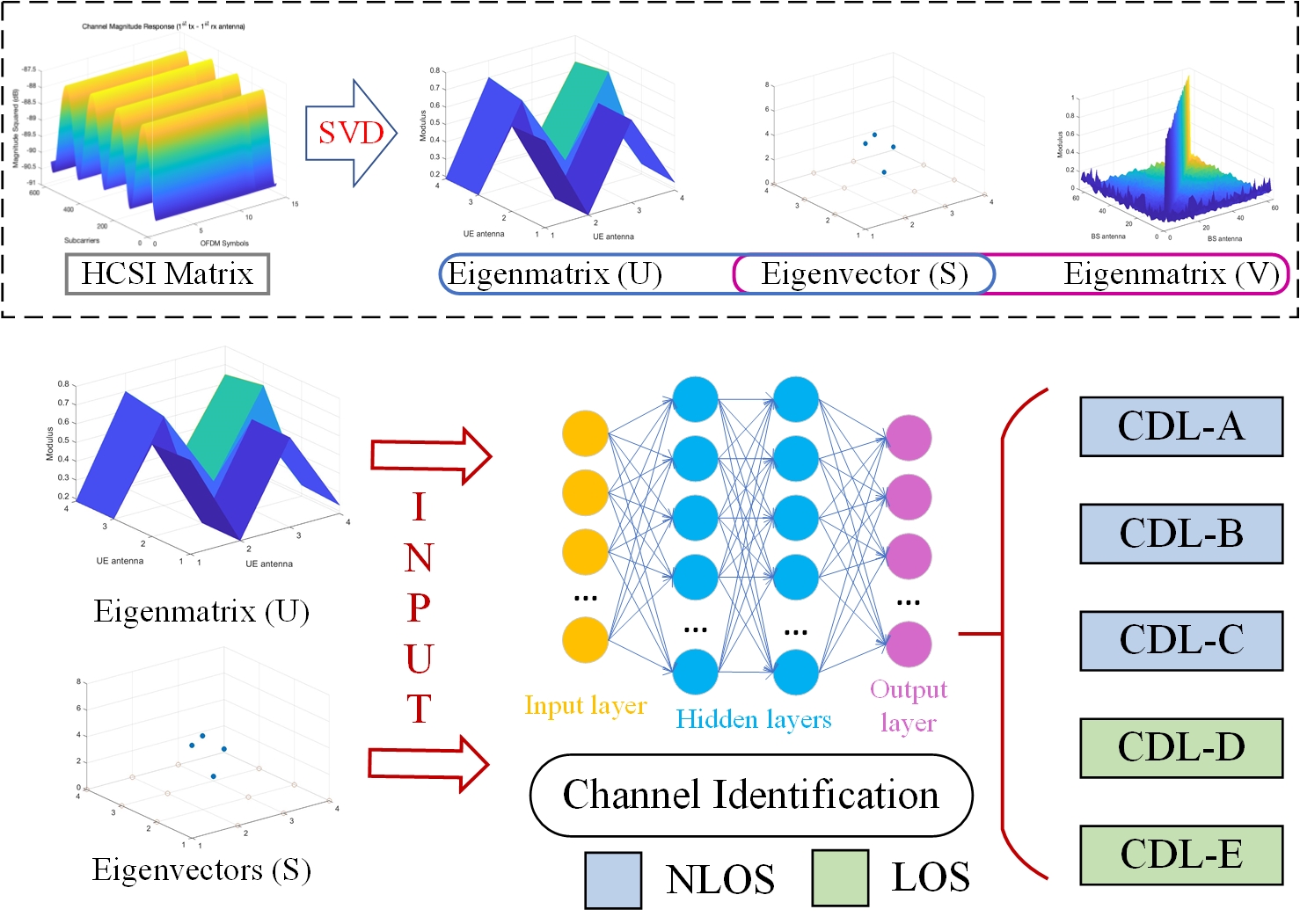}
	\caption{Illustration of the architecture of proposed novel EMEV based channel identification.}
	\label{fig:zongtu}
\end{figure}
\subsection{EMEV based Channel Identification Architecture}
This paper proposes a novel channel identification architecture based on eigenmatrix and eigenvector. Instead of the whole CSI matrix $\mathbf{H}$, its eigenvector  $\mathbf{S}$ and eigenmatrix  $\mathbf{U}$ will be used as the input object for identification neural network. $\mathbf{U}$ and  $\mathbf{S}$ of $\mathbf{H}$ can be obtained by SVD transformation. The proposed channel identification architecture is shown in Fig. \ref{fig:zongtu}, namely eigenmatrix and eigenvector (\textbf{EMEV}) based channel identification architecture.

Due to frequent movement of UEs relative to the BS,  UE should feedback the channel type information to  BS frequently in the frequency division duplexing (FDD) communication system. This paper focuses on the CDL channel, so we make the assumption that the CSI matrix $\mathbf{H}\in \mathbb{C}^{N_{RB} \times N_r \times N_t}$. The $N_{RB}$ denotes the number of resource block (RB), $N_t$ and $N_r$ are the number of BS and UE antennas ($N_t \gg N_r$).   As is shown in Fig. \ref{fig:zongtu}, the CSI matrix $\mathbf{H}$ will be divided into three parts after SVD transformation, namely eigenmatrix $\mathbf{U} \in \mathbb{C}^{N_{RB}  \times N_r \times N_r}$, eigenvector $\mathbf{S} \in \mathbb{R}^{N_{RB} \times N_r}$ and eigenmatrix $\mathbf{V} \in \mathbb{C}^{N_{RB}  \times N_t \times N_t}$. And then  $\mathbf{U}$ and  $\mathbf{S}$ will be input into the neural network, which is called EMEV based channel identification network (\textbf{EMEV-IdNet}). Compared with the size of CSI matrix $\mathbf{H}$, the size of  $\mathbf{U}$ and  $\mathbf{S}$ are much smaller, i.e.  EMEV-IdNet adopts a lightweight design.  After the trained EMEV-IdNet, we can identify accurately the channel type among CDL-A to CDL-E.  The EMEV based channel identification hopes to eliminate the redundant information in  CSI matrix $\mathbf{H}$ through the  SVD transformation, further make the neural network lightweight as much as possible.

\begin{table*}[htbp]
	\centering
	\caption{The hyper-parameters setting  and analysis of parameters and FLOPs. }
	\begin{tabular}{cccccc}
		\toprule
		Layer name & Hyper-parameters & Activation & Output shape & Parameter size & FLOPs \\ \hline
		Input$(\mathbf{U})$     &          --        &        --    &    $N_{RB} \times N_r \times N_r \times 2$       &      --    &    --   \\
		Input$(\mathbf{S})$    &          --        &      --      &     $N_{RB} \times N_r \times 1$       &      --            &   --    \\
		Conv3D\_1$(\mathbf{U})$ &  \multirow{2}{*}{\begin{tabular}[c]{@{}c@{}}Filter = 16,\\ Kernel = 3.\end{tabular}}  &  Leaky Relu  &  $N_{RB} \times N_r \times N_r \times 16$ &  $16 \times 2 \times 3^2$  & $(N_{RB} \times N_r \times N_r \times 16) \times (2 \times 3^2)$    \\
		Conv2D\_1$(\mathbf{S})$ &      &   Leaky Relu  &    $N_{RB} \times N_r  \times 16$  & $16 \times 2 \times 3^2$   &  $(N_{RB} \times N_r \times 16) \times (2 \times 3^2)$      \\
		Conv3D\_2$(\mathbf{U})$  &  \multirow{2}{*}{\begin{tabular}[c]{@{}c@{}}Filter = 32,\\ Kernel = 3.\end{tabular}}   &  Leaky Relu  &   $N_{RB} \times N_r \times N_r \times 32$  &    $32 \times 16 \times 3^2$  &     $(N_{RB} \times N_r \times N_r \times 32) \times (2 \times 3^2)$     \\
		Conv2D\_2$(\mathbf{S})$  &       &    Leaky Relu &   $N_{RB} \times N_r  \times 32$  &  $32 \times 16 \times 3^2$&   $(N_{RB} \times N_r \times 32) \times (2 \times 3^2)$     \\
		Conv3D\_3$(\mathbf{U})$ &  \multirow{2}{*}{\begin{tabular}[c]{@{}c@{}}Filter = 16,\\ Kernel = 3.\end{tabular}}   &  Leaky Relu  &  $N_{RB} \times N_r \times N_r \times 16$ &    $16 \times 32 \times 3^2$ &    $(N_{RB} \times N_r \times N_r \times 16) \times (2 \times 3^2)$      \\
		Conv2D\_3$(\mathbf{S})$ &    &  Leaky Relu   &    $N_{RB} \times N_r  \times 16$&  $16 \times 32 \times 3^2$   &   $(N_{RB} \times N_r \times 16) \times (2 \times 3^2)$     \\
		Concatenate&    --      &    --  &   $N_{RB}  N_r (N_r +1) \times 16 $ &    0 &   0  \\
		FCLayer\_1&   Units = 128    &    Relu    &   $128 \times 1$   & \scriptsize   $[N_{RB} N_r  (N_r +1) \times 16] \times 128 $  &  \scriptsize  $2 \times [N_{RB}  N_r (N_r +1) \times 16] \times 128 $       \\
		FCLayer\_2&   Units = 32 &   Relu  &   $32 \times 1$  &    $128 \times 32$   & $2 \times 128 \times 32$    \\
		FCLayer\_3&     Units = 5    & Softmax  &    $5 \times 1 $ &   $32 \times 5$  &   $2\times 32 \times 5$    \\
		\bottomrule   
	\end{tabular}
	\label{table1}
\end{table*}
\subsection{Analysis of Algorithm}
\begin{algorithm}
	\caption{The algorithm of the proposed EMEV based channel identification scheme.}
	\label{emevalg}
	\KwIn{$\mathbf{H} \in \mathbb{C}^{N_{RB} \times N_r \times N_t} \gets$  CSI matrix; \\
		$\boldsymbol{y} \in \mathbb{R}^{5 \times 1} \gets$ label of channel type; \\
		$\eta \gets$ learning rate; $N_{epoch} \gets$ Number of epoches}	
	\KwOut{channel type: $\hat{\boldsymbol{y}} = \left\{ 0, 1, 2, 3, 4 \right\}$}
	\textbf{SVD transformation :} \\
	{Initialize $ \mathbf{U} \in \mathbb{C}^{N_{RB} \times N_r \times N_r}, \mathbf{S} \in \mathbb{R}^{N_{RB} \times N_r}$};\\
	\For{$i = 1, \cdots N_{RB}$ }{
		$\mathbf{U_t}, \boldsymbol{S_t}, \mathbf{V_t} = f_{svd}(\mathbf{H}(i, :, :)) $ \\
		\If{$\mathbf{U_t} \cdot \boldsymbol{S_t} \cdot \mathbf{V_t} == \mathbf{H}(i, :, :)$}{
			$\mathbf{U}(i, ;, :) = \mathbf{U_t}$ \\
			$\mathbf{S}(i, :) = \boldsymbol{S_t}$
		}
	}
	{Save $\mathbf{U} \in \mathbb{C}^{N_{RB} \times N_r \times N_r}, \mathbf{S} \in \mathbb{R}^{N_{RB} \times N_r}$.}\\
	\textbf{Training EMEV-IdNet :}\\
	{Load $\mathbf{U} \in \mathbb{C}^{N_{RB} \times N_r \times N_r}, \mathbf{S} \in \mathbb{R}^{N_{RB} \times N_r}, \boldsymbol{y} \in \mathbb{R}^{5 \times 1}$};\\
	{Initialize $\Omega_{3D}, \Omega_{2D}, \Omega_{FC}, Adam(\cdot)$}; \\
	\For{$t = 1, \cdots, N_{epoch}$}{
		$\hat{\boldsymbol{y}} = f_{net}(\mathbf{U}, \Omega_{3D}, \mathbf{S}, \Omega_{2D}, \Omega_{FC})$\\
		$loss_{t} =-\sum_{k=1} \boldsymbol{y}_{k} \cdot \log(\hat{\boldsymbol{y}_{k}})$ \\
		\If{$loss_{t}$ converges}{break}
		$(\Omega_{3D},\Omega_{2D},\Omega_{FC}) \gets Adam(\Omega, \eta,  \nabla loss_{t})$ \\
	}
	{Save $\Omega_{3D}, \Omega_{2D}, \Omega_{FC}, f_{net}(\cdot)$.}
\end{algorithm}
Although the proposed EMEV based channel identification pays more attention to unitary matrix $\mathbf{U}$ and eigenvector $\mathbf{S}$, the most common information for UEs is CSI matrix $\mathbf{H}$.  Therefore,  we consider CSI matrix $\mathbf{H}$ as the input object in \textbf{Algorithm \ref{emevalg}}.  $\mathbf{U}$ and $\mathbf{S}$  are  obtained after data preprocessing. As for the neural network, we construct a dual channel parallel model which  takes eigenmatrix  $\mathbf{U}$  and eigenvector $\mathbf{S}$ as input layer.

The EMEV based channel identification is shown in \textbf{Algorithm \ref{emevalg}}.
In  SVD transformation part, we consider the channel independence between different RBs. Therefore, we do SVD transformation on each RB, and finally concatenate eigenmatrx and eigenvector. In EMEV-IdNet training part, $\Omega_{3D}, \Omega_{2D}, \Omega_{FC}$ represent the weights of 3D convolution layers, 2D convolution layers and fully-connected  layers, respectively, and $f_{net}(\cdot)$ denotes the framework function of EMEV-IdNet. Meanwhile, Adam optimizer and categorical-crossentropy loss function  are used to improve the performance and convergence rate. The algorithm training goal is to minimize the loss function between the true channel label $\boldsymbol{y}$  and the output label $\hat{\boldsymbol{y}}$  by updating the weights as,
\begin{equation}
	\label{loss}
	(\Omega_{3D}^*, \Omega_{2D}^*,\Omega_{FC}^*) = \arg \mathop{\min}_{\Omega} -\sum_{k=1} \boldsymbol{y}_{k} \cdot \log(\hat{\boldsymbol{y}_{k}})
\end{equation}
where $k$ denotes the length of channel label $\boldsymbol{y}$.

\subsection{Structure of EMEV-IdNet}
\begin{figure}[htbp]
	\centering
	\includegraphics[width=2.8 in,trim=0 0 0 0,clip]{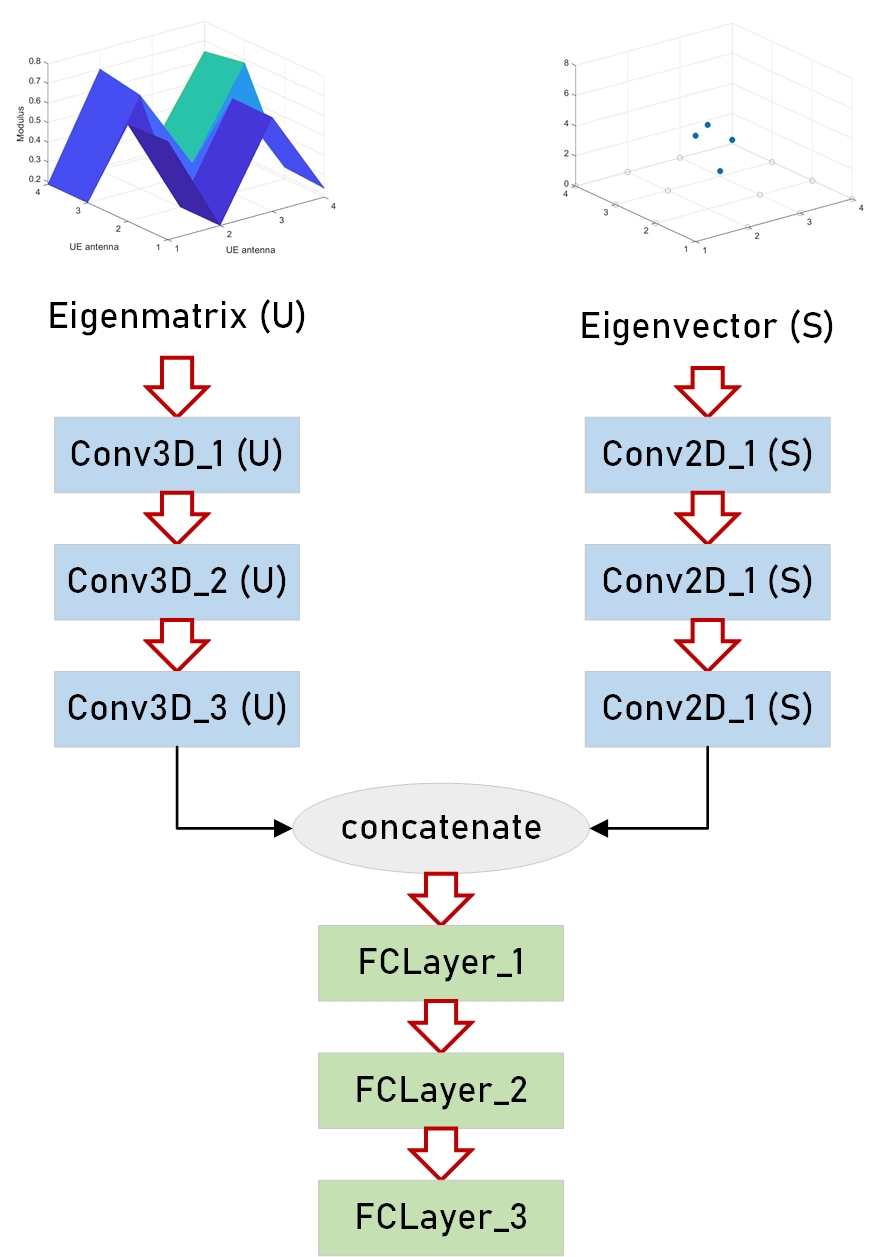}
	\caption{Illustration of the structure of EMEV-IdNet.}
	\label{fig:network}
\end{figure}

Fig. \ref{fig:network} shows the structure of the neural network used in our proposed scheme. The network we used in this paper is modifid from CNN. As is shown in figure, three dimensional (Conv3D) and two dimensional  convolutional layers (Conv2D) are used to extract special features for  $\mathbf{U}$ and $\mathbf{S}$, respectively. After the two parallel feature extraction blocks, an un-trainable concatenate layer is used. Then the two output high dimensional feature maps of $\mathbf{U}$ and $\mathbf{S}$ are concatenated. Finally, we rely on three fully-connected layers (FCLayer) to complete the identification task.
Meanwhile, Tab. \ref{table1}  shows the hyper-parameters, activation and output shape of every layer in detail. Meanwhile, the analysis of parameters and floating point operations (FLOPs) are presented.

\section{Simulation and Experiments}
This section will introduce the simulation experiments carried out in this paper in detail, including the experimental platform, simulation data, experimental results and so on.  All the simulations and experiments are carried out on the workstation with CentOS 7.0. The workstation is equipped with 2 Intel(R) Xeon(R) Silver 4210R CPU and 4 Nvidia RTX 2080Ti GPU, it also has 192GB RAM. The dataset we used in this paper and the simulation codes  can be found at Github\footnote{Github link: https://github.com/CodeDwan/EMEV-channle-identification}.

\subsection{Dataset Generation}
With the help of Matlab 5G toolbox and communication toolbox, we define a standard CDL channel object and carry on the link-level simulation. The dataset used in our experiments are extracted from the link-level simulator. Tab. \ref{data} shows the alternative parameter and  default values in the data generator. Both UE and BS antennas use uniform panel array (UPA).
Note that 10,000 original data samples  are  generated for each CDL channel  (50,000 data samples in all). And we divide the training data, validation data and test data in the proportion of $65:15:20$.  Using  the  StratifiedShuffleSplit function provided by sckit-learn, we can ensure that the data samples of each label is uniformly distributed.
\begin{table}[htbp]
	\caption{Alternative parameter and  default settings in data generator}
\begin{tabular}{|c|ccccc|}
	\hline
	\multirow{2}{*}{\begin{tabular}[c]{@{}c@{}}Channel\\ Type\end{tabular}} & \multicolumn{3}{c|}{NLOS}                                                               & \multicolumn{2}{c|}{LOS}           \\ \cline{2-6}
	& \multicolumn{1}{c|}{CDL-A}  & \multicolumn{1}{c|}{CDL-B}  & \multicolumn{1}{c|}{CDL-C}  & \multicolumn{1}{c|}{CDL-D} & CDL-E \\ \hline
	NRB    & \multicolumn{5}{c|}{13}               \\ \hline
	Center Frequency     & \multicolumn{5}{c|}{28 GHz}                \\ \hline
	Subcarrier spacing   & \multicolumn{5}{c|}{60 KHz}                  \\ \hline
	UE Speed    & \multicolumn{5}{c|}{\{4.8, 24, 40, 60\} km/h}           \\ \hline
	Delay spreads          & \multicolumn{1}{c|}{129 ns} & \multicolumn{1}{c|}{634 ns} & \multicolumn{1}{c|}{634 ns} & \multicolumn{1}{c|}{65 ns} & 65ns  \\ \hline
	BS antenna          & \multicolumn{5}{c|}{UPA $[8. 8] = 64$}               \\ \hline
	UE antenna            & \multicolumn{5}{c|}{UPA $[2, 2] = 4$}                \\ \hline
\end{tabular}
\label{data}
\end{table}

\subsection{Experimental Results}
\begin{figure}[htbp]
	\centering
	\includegraphics[width=3.0 in,trim=0 0 0 0,clip]{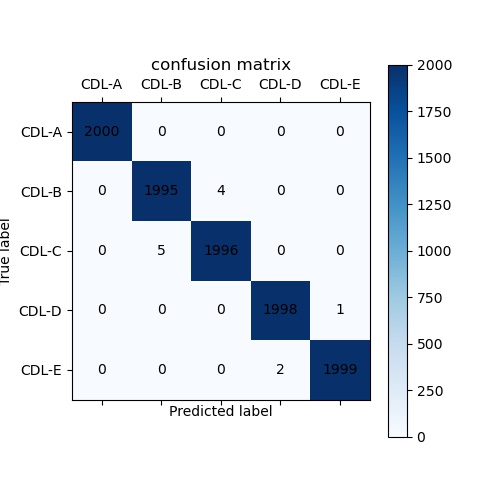}
	\caption{Illustration of the confusion matrix of EMEV-IdNet.}
	\label{fig:fenlei}
\end{figure}
In order to verify the feasibility and robustness of the scheme proposed in this paper, we design two experiments: one is to research the perfect CSI matrix, and the other is the imperfect CSI matrix with noise.  At the same time, in order to conduct comparative experiments, we design two different networks. One is a two channel identification network based on eigenmatrix and eigenvector proposed in this paper, named EMEV-IdNet. The other is  directly inputting CSI matrix $\mathbf{H}$ into the network for identification, named CSI-based identification network (CSI-IdNet) .

\begin{table}[htbp]\scriptsize
	\caption{The accuracy of EMEV-IdNet and CSI-IdNet with noise}
	\centering
	\begin{tabular}{ccccccc}
		\toprule
		SNR (dB)   & 10 & 12 & 14 & 16 & 18 & 20 \\ \hline
		EMEV-IdNet & 70.26\%   & 80.68\%   &  89.89\%  & 97.39\%   & 99.62\%   & 99.80\%   \\
		CSI-IdNet  &   87.55\% &  96.85\%  &  99.55\%  &  99.97\%  &  99.98\%  & 100.00\%   \\
		\bottomrule
	\end{tabular}
	\label{noise result}
\end{table}
\begin{figure}[htbp]
	\centering
	\includegraphics[width=3.5 in,trim=0 0 0 0,clip]{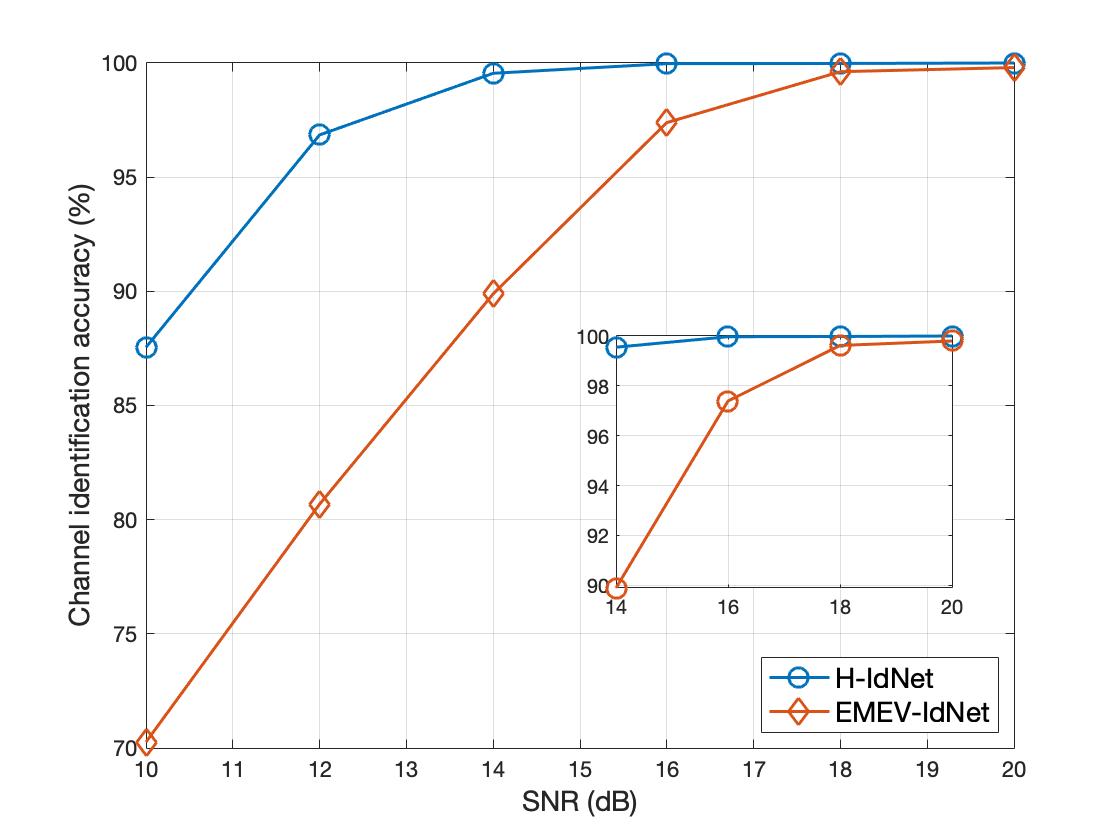}
	\caption{Accuracy comparison of EMEV-IdNet and CSI-IdNet with different noise.}
	\label{fig:noiseresult}
\end{figure}
First, we verify the feasibility of the proposed EMEV-IdNet.  The confusion matrix in Fig. \ref{fig:fenlei}  shows that CDL-A to CDL-C are  strictly separated from CDL
-D and CDL-E. That is to say, LOS and NLOS channels are perfectly recognized, and the CDL-A channel achieves perfect identification accuracy. What's more, the identification accuracy of CDL-B to CDL-E reaches $99.75\%, 99.80\%, 99.90\%$  and $ 99.95\%$ respectively. The comprehensive recognition accuracy is $99.88\%$
Secondly, it is difficult for UEs to obtain perfect CSI matrix in practical scenarios. Therefore, we conduct noise experiments and verify the robustness of  EMEV-IdNet. In this part, additive white Gaussian noise(AWGN) is used to simulate the imperfect CSI matrix, which can be expressed as,
\begin{equation}
	\begin{aligned}
		SNR (dB) & = 10 \cdot \log \left( \frac{P_{\mathbf{H}}}{P_{n}} \right) \\
		noise \sim N(0, P_{n} )& = \frac{1}{\sqrt{2\pi P_n}} \cdot \exp \left( -\frac{x^2}{2P_n} \right)
	\end{aligned}
\label{awgn}
\end{equation}
where $P_{\mathbf{H}}$ and $P_n$ are the respective power of CSI matrix $\mathbf{H}$ and $noise$, $N(0, P_n)$ represents normal  distribution with  $mean = 0$  and $variance =P_n$. We use the test dataset containing AWGN to test the perfect identification network (IdNet). The perfect IdNet means the network is trained with perfect CSI dataset. The test accuracy can be seen from Tab. \ref{noise result} and Fig. \ref{fig:noiseresult}. From the experimental results, the robustness of EMEV-IdNet is slightly worse than CSI-IdNet.  If we define the acceptable threshold of recognition accuracy is $acc \geq 95\%$, then the noise tolerated by EMEV-IdNet is SNR $=16$ dB, while the one CSI-IdNet can  tolerate SNR $=12$ dB.

\subsection{Analysis of Overhead}
In this part, we will analyse the overhead of EMEV-IdNet and CSI-IdNet, including the number of parameters, FLOPs, model size and response delay\footnote{The response delay is test on Nvidia RTX 2080Ti.}. The parameters and model size  measure space complexity and storage overhead. The FLOPs and response delay can measure time complexity and computing overhead.
As is shown in Tab. \ref{overhead}, the comprehensive overhead of EMEV-IdNet  is about only 10\% of CSI-IdNet. However, the performance of EMEV-IdNet can achieve almost 90\% comparing with CSI-IdNet. In summary, the proposed EMEV-IdNet  conforms to the lightweight design and ensures performance as much as possible while greatly reducing overhead.
\begin{table}[htbp]
	\centering
	\caption{Comparison of overhead between EMEV-IdNet and CSI-IdNet}
	\begin{tabular}{|c|c|c|}
		\hline
		& EMEV-IdNet & CSI-IdNet \\ \hline
		Parameters     &     575,157       &    6,848,869 \\ \hline
		FLOPs          &     14 M       &   204 M        \\ \hline
		Model size     &       7.1 MB     &       82.3 MB    \\ \hline
		Response delay &      88.23 us      &    294.29 us       \\ \hline
	\end{tabular}
	\label{overhead}
\end{table}

\section{Conclusion}
In this paper, we proposed a novel EMEV based channel identification architecture with lightweight design. In this paper, SVD transformation was adopted as data preprocessing. The obtained eigenmatrix and eigenvector can be used not only for channel identification, but also for calculation of precoding matrix.  It was confirmed that EMEV-IdNet proposed in this paper has a great identification performance and very slim overhead.  When using  perfect CSI matrix, EMEV-IdNet can achieve a comprehensive identification accuracy of $99.88\%$, in which the CDL-A channel can be perfectly recognized. When  AWGN is introduced to simulate the actual channel estimation error, EMEV-IdNet can tolerate the noise with SNR $=16$ dB if the threshold is  $acc \geq 95\%$. At the same time, we analyzed the comprehensive overhead of EMEV-IdNet, which is more suitable for deployment at UE.  The EMEV based channel identification architecture proposed in this paper meets the needs of mmWave communication. Considering the massive number of edge devices in the IoE network, the proposed architecture is lightweight and effective for edge computing deployment.  It can be widely applied to  edge equipment in mmWave systems to achieve the high quality of service (QoS) of IoE communication.

\section{Acknowledgements}
This work is supported by the open research fund of the Key Laboratory of Dynamic Cognitive System of Electromagnetic Spectrum Space, Ministry of Industry and Information Technology under grant KF20202106 and National Natural Science Foundation of China under Grant 61901228.

\bibliographystyle{IEEEtran}
\bibliography{channel_id}

\begin{thebibliography}{10}
\providecommand{\url}[1]{#1}
\csname url@samestyle\endcsname
\providecommand{\newblock}{\relax}
\providecommand{\bibinfo}[2]{#2}
\providecommand{\BIBentrySTDinterwordspacing}{\spaceskip=0pt\relax}
\providecommand{\BIBentryALTinterwordstretchfactor}{4}
\providecommand{\BIBentryALTinterwordspacing}{\spaceskip=\fontdimen2\font plus
\BIBentryALTinterwordstretchfactor\fontdimen3\font minus
  \fontdimen4\font\relax}
\providecommand{\BIBforeignlanguage}[2]{{%
\expandafter\ifx\csname l@#1\endcsname\relax
\typeout{** WARNING: IEEEtran.bst: No hyphenation pattern has been}%
\typeout{** loaded for the language `#1'. Using the pattern for}%
\typeout{** the default language instead.}%
\else
\language=\csname l@#1\endcsname
\fi
#2}}
\providecommand{\BIBdecl}{\relax}
\BIBdecl

\bibitem{LiYe_IoT}
A.~Froytlog, T.~Foss, O.~Bakker, G.~Jevne, M.~A. Haglund, F.~Y. Li, J.~Oller,
  and G.~Y. Li, ``{Ultra-Low Power Wake-up Radio for 5G IoT},'' \emph{IEEE
  Communications Magazine}, vol.~57, no.~3, pp. 111--117, 2019.

\bibitem{Kato_6G1}
B.~Mao, F.~Tang, Y.~Kawamoto, and N.~Kato, ``{AI Models for Green
  Communications Towards 6G},'' \emph{IEEE Communications Surveys \&
  Tutorials}, vol.~24, no.~1, pp. 210--247, 2022.

\bibitem{survey1}
O.~E. Ayach, S.~Rajagopal, S.~Abu-Surra, Z.~Pi, and R.~W. Heath, ``{Spatially
  Sparse Precoding in Millimeter Wave MIMO Systems},'' \emph{IEEE Transactions
  on Wireless Communications}, vol.~13, no.~3, pp. 1499--1513, 2014.

\bibitem{survey2}
I.~A. Hemadeh, K.~Satyanarayana, M.~El-Hajjar, and L.~Hanzo, ``{Millimeter-Wave
  Communications: Physical Channel Models, Design Considerations, Antenna
  Constructions, and Link-Budget},'' \emph{IEEE Communications Surveys \&
  Tutorials}, vol.~20, no.~2, pp. 870--913, 2018.

\bibitem{LiYe2019}
Z.~Qin, H.~Ye, G.~Y. Li, and B.-H.~F. Juang, ``{Deep Learning in Physical Layer
  Communications},'' \emph{IEEE Wireless Communications}, vol.~26, no.~2, pp.
  93--99, 2019.

\bibitem{Kato_6G2}
H.~Guo, J.~Li, J.~Liu, N.~Tian, and N.~Kato, ``{A Survey on
  Space-Air-Ground-Sea Integrated Network Security in 6G},'' \emph{IEEE
  Communications Surveys \& Tutorials}, vol.~24, no.~1, pp. 53--87, 2022.

\bibitem{Guojiajia}
J.~Guo, C.-K. Wen, and S.~Jin, ``{Deep Learning-Based CSI Feedback for
  Beamforming in Single- and Multi-Cell Massive MIMO Systems},'' \emph{IEEE
  Journal on Selected Areas in Communications}, vol.~39, no.~7, pp. 1872--1884,
  2021.

\bibitem{Mantoro2017}
T.~Mantoro, M.~A. Ayu, and M.~R. Nugroho, ``{NLOS and LOS of the 28 GHz Bands
  Millimeter-wave in 5G Cellular Networks},'' in \emph{2017 International
  Conference on Computing, Engineering, and Design (ICCED)}, 2017, pp. 1--5.

\bibitem{Zhang2013}
J.~Zhang, J.~Salmi, and E.~S. Lohan, ``{Analysis of Kurtosis-based LOS/NLOS
  Identification Using Indoor MIMO Channel Measurement},'' \emph{IEEE
  Transactions on Vehicular Technology}, vol.~62, pp. 2871--2874, 2013.

\bibitem{Huang2017}
C.~Huang and X.~Zhang, ``{LOS-NLOS Identification Algorithm for Indoor Visible
  Light Positioning System},'' in \emph{2017 20th International Symposium on
  Wireless Personal Multimedia Communications (WPMC)}, 2017, pp. 575--578.

\bibitem{Diamant2014}
R.~Diamant, H.~P. Tan, and L.~Lampe, ``{LOS and NLOS Classification for
  Underwater Acoustic Localization},'' \emph{IEEE Transactions on Mobile
  Computing}, vol.~13, pp. 311--323, 2 2014.

\bibitem{Xiao2017}
F.~Xiao, Z.~Guo, H.~Zhu, X.~Xie, and R.~Wang, ``{AmpN: Real-time LOS/NLOS
  Identification with WiFi},'' in \emph{2017 IEEE International Conference on
  Communications (ICC)}, 2017, pp. 1--7.

\bibitem{Zeng2018}
T.~Zeng, Y.~Chang, Q.~Zhang, M.~Hu, and J.~Li, ``{CNN-Based LOS/NLOS
  Identification in 3-D Massive MIMO Systems},'' \emph{IEEE Communications
  Letters}, vol.~22, pp. 2491--2494, 12 2018.

\bibitem{Zhang2019}
S.~Zhang, C.~Yang, D.~Jiang, X.~Kui, S.~Guo, A.~Y. Zomaya, and J.~Wang,
  ``{Nothing Blocks me: Precise and Real-time LOS/NLOS Path Recognition in RFID
  Systems},'' \emph{IEEE Internet of Things Journal}, vol.~6, pp. 5814--5824, 6
  2019.

\bibitem{Huang2020}
C.~Huang, A.~F. Molisch, R.~He, R.~Wang, P.~Tang, B.~Ai, and Z.~Zhong,
  ``{Machine Learning-Enabled LOS/NLOS Identification for MIMO Systems in
  Dynamic Environments},'' \emph{IEEE Transactions on Wireless Communications},
  vol.~19, pp. 3643--3657, 6 2020.

\bibitem{Jiang2020}
C.~Jiang, S.~Chen, Y.~Chen, D.~Liu, and Y.~Bo, ``{An UWB Channel Impulse
  Response De-Noising Method for NLOS/LOS Classification Boosting},''
  \emph{IEEE Communications Letters}, vol.~24, no.~11, pp. 2513--2517, 2020.

\bibitem{Cui2021}
Z.~Cui, Y.~Gao, J.~Hu, S.~Tian, and J.~Cheng, ``{LOS/NLOS Identification for
  Indoor UWB Positioning Based on Morlet Wavelet Transform and Convolutional
  Neural Networks},'' \emph{IEEE Communications Letters}, vol.~25, no.~3, pp.
  879--882, 2021.

\bibitem{Zheng2020}
Q.~Zheng, R.~He, B.~Ai, C.~Huang, W.~Chen, Z.~Zhong, and H.~Zhang, ``{Channel
  Non-Line-of-Sight Identification Based on Convolutional Neural Networks},''
  \emph{IEEE Wireless Communications Letters}, vol.~9, no.~9, pp. 1500--1504,
  2020.

\bibitem{3GPP38901}
3GPP, ``{Study on channel model for frequencies from 0.5 to 100 GHz (Relase
  16)},'' {3rd Generation Partnership Project (3GPP)}, Technical Report (TR)
  38.901, 12 2019, version 16.1.0.

\bibitem{3GPP36873}
------, ``{Study 3D Channel Model for LTE (Relase 12)},'' {3rd Generation
  Partnership Project (3GPP)}, Technical Specification (TS) 36.873, 1 2018,
  version 14.1.0.

\end{thebibliography}

\end{document}